# Evaluation of Ultra Low Dose chest CT imaging for Covid 19 diagnosis and follow up

**Fariba Zarei[1], Reza Jalli[1], Sabyasachi Chatterjee[2], Mehrzad Lotfi[1], Pooya Iranpour[1], Vani VC[3], Sedigheh Emadi[1], Rezvan Ravanfar Haghighi[1,*]**

[1] Medical Imaging Research Center, Shiraz University of Medical Sciences, Shiraz 7193635899, Iran

[2] Ongil, 79 D3, Sivaya Nagar, Reddiyur Alagapuram, Salem 636004. India

[3] Department of Instrumentation and Applied Physics, Indian Institute of Science, Bangalore 560012. INDIA

Corresponding author: Rezvan Ravanfar Haghighi[1,*]

**Abstract:**

**Objective:** Computed Tomography (CT) has an important role to detect lung lesion related to Covide 19. The purpose of this work is to obtain diagnostic findings of Ultra-Low Dose (ULD) chest CT image and compare with routine dose chest CT.

**Material and Methods:** Patients, suspected of Covid 19 infection, were scanned successively with routine dose, and ULD, with 98% or 94% dose reduction, protocols. Axial images of routine and ULD chest CT were evaluated objectively by two expert radiologists and quantitatively by Signal to Noise Ratio (SNR) and pixel by pixel noise measurement.

**Results:** It was observed that the ULD and routine dose chest CT images could detect Covid 19 related lung lesions in patients with PCR positive test. Also, SNR and pixel noise values were comparable in these protocols.

**Conclusion:** ULD chest CT with 98% dose reduction can be used in non-pandemic situation as a substitute for chest radiograph for screening and follow up. Routine chest CT protocol can be replaced by ULD, with 94% dose reduction, to detect patients suspected with Covid 19 at an early stage and for its follow up.

**Advances in knowledge:** ULD chest CT with 94% dose reduction is able to detect lung lesions related to Covid 19. ULD chest CT is a suitable substitute for routine dose chest CT especially in pediatric, young and pregnant patients suspected with Covid 19.

**Introduction:**

Since the first report of Covid 19 affliction in December 2019, the epidemic has spread to several countries affecting tens of millions of people, has caused over hundred thousand deaths and threatens to spread further in the first phase, with risk of resurgent return in future. As a matter of public health policy and urgency of treatment, it is important to arrive at a protocol for quick confirmation at an early stage and also for monitoring the prognosis by repeated tests in future. Some of the studies have reported that un-enhanced chest CT reveals lung lesions related to Covid 19. Chest CT can thus be considered as an important diagnostic tool, comparable to the Real-Time Polymerase Chain Reaction (RT-PCR) which is known as a gold standard test to detect new Coronavirus or Covid 19 affected patients [1,2].

Diagnostics by CT, though known for its precise detection ability has the disadvantage of high dose of x-ray exposure to the patient, with the risk of creating several stochastic effects including onset of carcinoma in future. Literature review demonstrated that different types of lung lesions can be detected in chest x-ray with less than 1 mSv effective dose (more than 90% dose reduction compare to routine or standard dose), which is very close to both, Posterior-Anterior (PA) and Lateral chest radiograph, when iterative reconstruction is available. This CT protocol is known as Ultra-Low Dose or ULDCT. It is shown in this study that the ULDCT chest image, due to its three dimensional nature, is capable of detecting normal and abnormal structures better than by plain radiograph [3-6]. In resent publication, Dangis A. et al [7] have shown that low dose chest CT with less than 1 mSv can be used to diagnose lung lesion related to Covid 19 in emergency ward.

In developing countries like ours, prolonged testing process like the PCR are not easily accessible. Thus chest CT has a valuable role in early detection of lung lesion related to Covid 19. Further, radiation dose reduction by using ULD chest CT protocols can have important role in radiation protection of the patient in our hospitals.

In this rapid communication, we particularly examine the importance of ultra-low dose computed tomography (ULDCT) of the chest, to detect lung lesions related to Covid 19, in its early stage and follow up. Physical explanations are used to justify the finding of this study.

**Materials and methods:**

This study was approved by the national ethical committee in biomedical research (ethics ID: 99-01-48-22204-196109). Patients with signs of Covid 19 infection (respiratory problem) were told about the increased radiation dose that they would receive from an additional ULDCT test. They volunteered themselves for this test and the ULDCT test was performed only after they signed the consent form. Patients were scanned successively with routine and ULD chest CT protocols, without patient motion between two scan protocols. The 128-MDCT Philips Ingenuity system was used to scan the patients. Routine dose protocol uses 120 kVp along with the tube current modulation to scan the chest of the patients referred to the CT department, being done routinely by default. Two ULD scanning parameters were used to scan the volunteer patient, first group of patients were scanned by 98% dose reduction protocol (compare to routine dose) and the second group was scanned by about 94% dose reduction. In the two ULD chest CT protocols, 80 kVp was selected as the excitation tube voltage, while the tube current time product was fixed at 10 and 25 mAs for the first and second group of patients respectively.

It has to be mentioned that the patients with underlying diseases (such as cancer or cardiovascular disorders), sever respiratory distress (poor condition) and those who did not agreed to do the ULD chest CT were excluded from this study.

The other scan parameters such as, slice thickness and intervals (2mm and 5mm), gantry rotation time (0.4 sec) and so on, were the same in both routine and ULD CT protocols. Routine CT images (axial and coronal) were reconstructed by the hybrid iterative reconstruction software iDose level 4 while ULD CT images (axial and coronal) were reconstructed by complete iterative reconstruction IMR (Iterative Model Reconstruction) level 1. Dose indices such as volume CT dose index ($CTDI_{vol}$ in mGy) and dose length product (DLP in mGy.cm) were recorded for dose assessment purpose. Effective dose was calculated by the product of conversion factor (mSv/mGy.cm) and DLP [8]. Axial and coronal slices were reconstructed with iDose level 4 and IMR level 1 for routine and ULD CT respectively.

Signal to Noise Ratio (SNR) was measured at the air of trachea and descending aorta for routine and ULD chest CT images. This was done by noting the mean HU values and the standard deviation in HU, as recoded on the ROI from trachea and aorta without contamination with neighboring structures. SNR measurement was used as a quantitative method for the evaluation of image quality.

The images are displayed side by side in a diagnostic monitor (Barco monitor, MDMC- 12133, Belgium). They were examined independently by two radiologists, each with 15 years' experience and were blinded to each other's report. It has to be mentioned that the chest CT images which were included in this study, were those that showed the patient to be affected by Covid 19, as confirmed by PCR-test.

**Results:**

Objective evaluation of the chest CT images in Covid 19 affected patients by two expert radiologists revealed that both ULD chest CT (98% and 94% dose reductions) protocols can detect corona virus related lung lesions which were diagnosed in routine dose CT protocol. Figures (1) and (2) are typical images from two patients for whom PCR test had confirmed Covid 19 infection.

Although, figure (1.a) was taken by 98% dose reduction, it was able to detect Atoll sign (reverse halo sign) as clearly as in routine dose protocol, given in Fig. (1.b) in chest CT images of Covid 19 patients (confirmed by PCR test). It has to be mentioned that the effective dose to this patient (calculated by the product of 0.014 into DLP) were 6.6 and 0.09 mSv in routine and ultra-low dose protocols respectively.

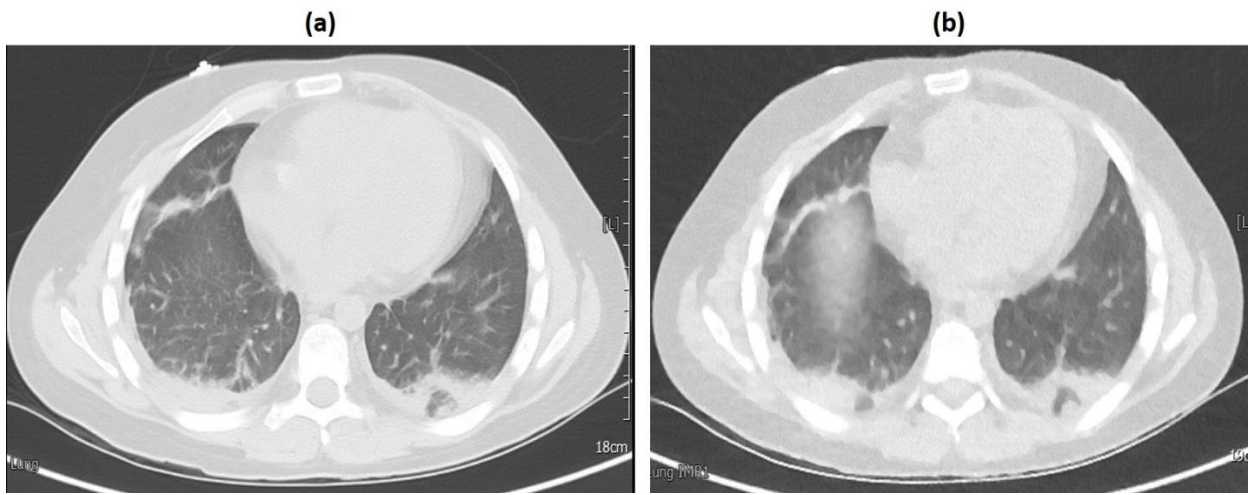

*FIG. 1. Axial chest CT images show the Atoll sign or reverse halo sign, in a 21 years' female Covid 19 patient, scanned by (a) routine dose (6.6 mSv effective dose) and (b) ultra-low dose (98% dose reduction with 0.09 mSv effective dose) protocols.*

It can be seen in Fig. 2 (a-d) that the chest CT images of a Covid 19 infected patient, scanned at routine dose, the effective dose is about 3.7 mSv, (Fig. 2.a and c) and for 94% dose reduction, the effective dose is about 0.22 mSv, (Fig. 2.b and d). Axial slices of chest CT with ULD (94% reduced dose) protocol reveals Crazy Paving appearance, although its resolution is less than that of the routine dose image (Fig. 2.b). Ground glass opacity (GGO) and consolidation which are seen in routine dose axial CT image (Fig 2.c) of the same patient, can be detected in ULD CT image (Fig. 2.d) with about 94% dose reduction equally clearly by visual discrimination by the radiologist.

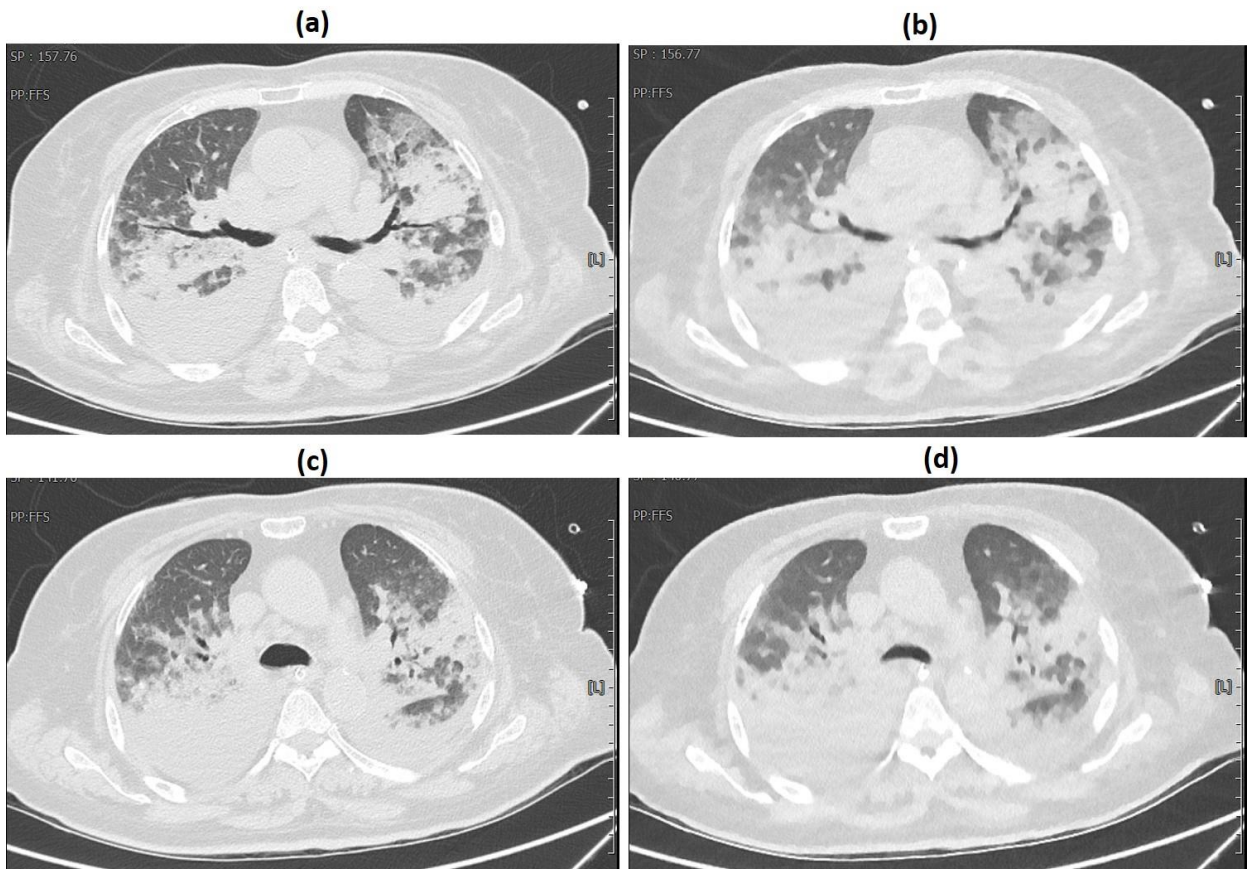

*FIG.2. Axial chest CT images of 71 year-old female patient infected with Covid 19 (PCR positive) shows crazy paving pattern in routine dose (a) and 94% dose reduction or ULD (b) protocols. Ground Glass Opacities (GGO) can be seen in routine dose (c) and ULD (d) protocols. Effective doses of this patient in routine and ULD (94% dose reduction) protocols were 3.7 mSv and 0.22 mSv respectively.*

**Quantitative image quality comparison:** The SNR, were calculated for the mean HU and in the pixel by pixel noise, in blood and air inside the aorta and trachea, were taken for the ULD (with 98% and 94% dose reductions) and routine dose chest CT images. These estimates revealed that the SNR of the ULD (with 98% and 94% dose reductions) and routine dose chest CT images, are comparable to each other. SNR and pixel by pixel noise for ULD and routine dose chest CT images are presented in Table I.

*Table I. Signal to Noise Ratio and pixel noise were measured at the descending aorta and trachea for axial chest CT images of the patients suspected to Covid 19 for routine dose and ultra-low dose (98% and 94% dose reduction) protocols.*

| Chest CT Protocol | SNR | | Pixel by pixel Noise | |
|---|---|---|---|---|
| | Aorta | Trachea | Aorta | Trachea |
| Routine | 360 | 6 | 33 | 0.7 |
| ULD (98% dose reduction) | 352 | 5.2 | 33 | 0.64 |
| ULD (94% dose reduction) | 316 | 4 | 29 | 0.55 |

The results of CT dose measurements show that CTDI$_{vol}$ of ULDCT protocols are fixed at 0.2 and 0.5 mGy, and their DLPs are 7.4 (6.4-9.5) mGy.cm and 18.35 (15.6-20.2) mGy.cm for 98% and 94% dose reduction protocols. The mean of CTDI$_{vol}$ and DLP in routine dose chest CT of Covid 19 patients are about 11 (6.6-20) mGy and 400mGy.cm (255-700 mGy.cm) respectively.

Effective dose of routine dose chest CT in Covid 19 affected patients is 5.7 mSv which is 55 and 22 times higher than that of ULD CT protocols with 98% (mean effective dose is equal to 0.1036) and 94% (mean effective dose is equal to 0.2569 mSv) respectively.

**Discussion:**

The ULD chest CT images with 98% and 94% dose reduction in 20 patients suspected of Covid 19, were examined by two expert radiologists blinded to each other report. Both of them independently reported the existence of the same features in both routine dose and ULD chest CT protocols. Some of the typical findings, as presented in the result section are; Ground glass opacity, consolidation, reticular lines, atoll or reverse halo sign and air bronchogram.

The authors of this study decided to use the 94% dose reduction protocol as a ULD CT protocol in patients suspected of Covid 19. Since chest CT images are used as a very important and valuable means to detect lung lesions related to new coronavirus (Covid 19) in early stage, ULD chest CT with 98% dose reduction can be used as a substitute to plain radiograph in non-pandemic situation.

This preliminary study shows the importance of ULDCT for early detection of lung infection in Covid 19 patients by identifying the abnormal structures in the lung such as ground glass opacities and consolidations. Two questions need to be addressed here, to be explained on the basis of CT's operational principles and thus justify the power of the method used. The CT machine works on the principle of x-ray attenuation coefficient of matter in the part of the radiation. The mean attenuation coefficient ($\hat{\mu}$) is given by $\hat{\mu} = [a(V)\rho_e + b(V) (\rho_e Z_{eff}^x )]$, where $\rho_e$ is the electron density and $Z_{eff}$ is the effective atomic number of the material with *x=3-4* being the exponent of the photoelectric absorption while *a(V)* and *b(V)* are the coefficients of the Compton scattering and photoelectric absorption. These quantities, *a(V)* and *b(V)* are averaged over the source spectrum of the CT, with *V* in *kVp*, is the operational voltage of the CT machine. While the Compton scattering part has a weak dependence on the photon energy*(E)*, the photoelectric absorption has a strong increase *(~1/$E^3$)* with decreasing *E*. Our computations with the well-known Boone-Seibert source spectrum [9] show that without filter, *a(80)=5.76×$10^{-25}$, b(80)=3.34×$10^{-27}$; a(120)=5.05×$10^{-25}$, b(120)=2.2×$10^{-27}$;* while on using a standard aluminum filter of *7 mm* thickness, we have*: a(80)=5.4×$10^{-25}$, b(80)=5.43×$10^{-28}$, a(120)=5.35×$10^{-25}$,b(120)= 4.75×$10^{-28}$,* where *a(V)* and *b(V)* are expressed in $cm^2$. Thus, in either case, the photoelectric effect contribution at *V=80 kVp* is much higher than that at *V=120 kVp*. This means that at *V=80 kVp,* any small difference in the effective atomic number $Z_{eff}$, gets highly magnified owing to the higher value of *b(80)* than *b(120)*and significant increase in $Z_{eff}^x$. This is the reason for more sensitive detectability of structures by increasing the photoelectric contribution (increasing contrast). Enhancement of the contribution from the Compton scattering part would be very limited, since *a(80)≈a(120)~1-0.90,* and the lung is made of low density material (Compton scattering is dominant) [10].

Thus, structures in the lung (any lesion with very small differences atomic number with air) could be detected in the ULD case even though these is a reduction in the SNR, in the primary stage, due the fall in the photon number (n), *(SNR ~$\sqrt{n}$)*. On using the Boone-Seibert formula, get the respective photon numbers to be *(a) n(80) =1.41×$10^6$, n(120)=3.40×$10^6$,* without added filter and

(b) $n(80) = 2.56 \times 10^3$ and $n(120)=3.25 \times 10^4$ for the case with aluminum added filter of 7mm thickness (the most common added filter in CT x-ray tube). The photon number for 80 kVp with filter is thus only 7.8% of the corresponding one in the 80kVp case while the SNR in the ULDCT case is thus lower by a factor of $\sqrt{2.56/3.25}$, i.e. 0.35 than that in the normal case. We believe, this SNR is still high enough for the iterative algorithm to use and bring the contrast, due to enhanced differences in photoelectric effect contributions [11].

Another important issue is that in the case of ULDCT the total number of photons is further reduced to $(1/50^{th})$ of that in the normal dose. This implies that even if multiple exposures are taken (less than 10, say) at follow-up, the total photon exposure would still remain a small fraction of what is given for the normal dose. The SNR, though reduced by a factor of 0.14, from that of the normal dose, is presumed to be not beyond the power of the Iterative Reconstruction of algorithm to correct for. This is important for the patients' safety as well as for that of the technical staff. This low voltage and mAs, used in ULDCT, would also protect the machines from breakdown, which is another important requirement for dealing with the situation under the threat of Covid 19 epidemic.

Since, the SNR and noise per pixel are comparable for chest CT images with routine, 98% and 94% dose reduction protocols, therefore, we can conclude that the noise increment due to dose reduction protocols are compensated by the iterative reconstruction (IMR level 1). Then chest CT images, taken by routine and ULD protocols, have adequate SNR and noise per pixel and do not deteriorate diagnostic image quality.

We have noticed while writing this paper that Dangis A. et al have reported that low dose low-dose submillisievert chest CT can be used for the diagnosis of COVID-19 [7]. We thus consider our paper to be extremely topical, more so because the dose in our observations is much lower (mean effective dose is 0.256 mSv) than that above authors (mean effective dose is 0.56 mSv) have used. Further, physical explanations of SNR and attenuation coefficient justified that structures in the image are not due to errors arising out of noise. Our work has been done with minimum dose that has been reported so far.

**Conclusion:**

Chest CT images with 94% dose reduction can be used to detect Covid 19 lesions at an early stage and be used as an important follow up tool, in the treatment. Reducing dose by ULD protocol reduces the stochastic effects such as genetic disorder and cancer in patients, especially in pediatric, young and pregnant patients. Also, using ULD CT protocols reduces the CT system's work load, therefore increases the stability and lifetime of x-ray tubes. In the background of the new pandemic (Covid 19), there is an increasing demand for chest CT requests in 2020 compared to the same season in the previous years. Its importance is in diagnosis of covid 19 patients is pointed out in the present paper. The authors of this study suggest that ULD chest CT protocol with 94% dose reduction should be used as a new guideline to substitute routine dose chest CT.

**No conflict of interest:** The authors have no conflict of interest.

**Acknowledgement:** The authors thank the Shiraz University of Medical Sciences for their encouragement and financial support of the project number 99-01-48-22204.